\documentclass[preprint]{aastex}

\shorttitle{Star Formation Near KR~140}
\shortauthors{Kerton et al.}

\begin{document}


\title{A Submillimeter View of Star Formation Near the \ion{H}{2} Region KR 140}

\author{C. R. Kerton}
\affil{Dominion Radio Astrophysical Observatory, Herzberg Institute of
Astrophysics,\\ National Research Council, P. O. Box 248, Penticton,
BC, V2A 6K3, Canada}
\email{charles.kerton@nrc.ca}

\author{P. G. Martin}
\affil{Canadian Institute for Theoretical Astrophysics and Department of
Astronomy, \\ University of Toronto, 60 St. George Street, Toronto, ON, M5S
3H8, Canada} 
\email{pgmartin@cita.utoronto.ca}

\author{D. Johnstone} 
\affil{Department of Astronomy, University of Toronto, \\ 60 St. George
Street, Toronto, ON, M5S 3H8, Canada}
\email{johnston@astro.utoronto.ca}

\and
\author{D. R. Ballantyne} 
\affil{Institute of Astronomy, University of Cambridge,\\ Madingley
Rd., Cambridge, CB3 0HA, UK}
\email{drb@ast.cam.ac.uk}


\begin{abstract}
We present the results of 450 and 850 $\mu$m continuum mapping of the
\ion{H}{2} region KR~140 using the SCUBA instrument on the JCMT. KR~140
is a small (5.7~pc diameter) \ion{H}{2} region at a distance of 
2.3$\pm$0.3 kpc. Five of the six IRAS point sources near KR~140 were mapped in
this study. Our analysis shows that two of these IRAS sources are
embedded late B type
stars lying well outside the \ion{H}{2} region, two are a part of
the dust shell surrounding the \ion{H}{2} region, and one is the
combined emission from an ensemble of smaller sources unresolved by IRAS.
We have discovered a number of relatively cold
submillimeter sources not visible in the IRAS data, ranging in size from
0.2 to 0.7~pc and in mass from 0.5 to 130~M$_\odot$.  The distribution
of masses for all sources is well characterized by a power law $N(>M)
\propto M^{-\alpha}$ with $\alpha = 0.5 \pm 0.04$, in agreement with the
typical mass function for clumped structures of this scale in molecular clouds.
Several of the submillimeter sources are found at the
\ion{H}{2}--molecular gas interface and have probably been formed 
as the result of the expansion of the \ion{H}{2} region.
Many of the submillimeter sources we detect are gravitationally bound
and most of these follow a mass--size relationship expected for objects
in virial equilibrium with non-thermal pressure support.  Upon the loss
of non-thermal support they could be sites of star formation.
Along with the two B stars that we have identified as possible cluster
members along with VES~735, we argue that five nearby highly-reddened
stars are in a pre-main-sequence stage of evolution.

\end{abstract}


\keywords{ \ion{H}{2}  Regions --- submillimeter --- stars: formation
--- ISM: clouds}


\section{INTRODUCTION}
\label{sec:intro}

KR~140 ($l=133\fdg 425$, $b=0\fdg055$; Kallas \& Reich 1980) is a
 small (5.7 pc diameter) \ion{H}{2} region located in the Perseus arm
 of the Galaxy at a distance
of 2.3$\pm$0.3 kpc (Kerton, Ballantyne, \& Martin 1999).  It is close to, but
apparently isolated from, the W3/4/5 star formation complex. The region
is ionized by VES~735, an
O8.5~V(e) star \citep{ker99}.  It has been the subject of a recent
multiwavelength study (Ballantyne, Kerton, \& Martin 2000) utilizing the data 
set of the Canadian Galactic Plane Survey (CGPS; Taylor 1999).

Surrounding the \ion{H}{2} region are six IRAS point sources (see
Table~\ref{tbl:iras} and Figure~\ref{fig:radioir}; there is a faint
source coincident with VES~735 as well).  These are the brightest
sources within several square degrees, suggestive of a region of
enhanced star formation.  In particular, infrared point sources located
at the periphery of \ion{H}{2} regions are of interest because they may
be associated with star formation induced by the expansion of the
\ion{H}{2} region (e.g., Sugitani, Fukui, \& Ogura 1991).
Characterization of these 
sources is important in order to form a complete picture of the star
formation history in that area.

Determining the true nature of these sources is hampered by the low
resolution of the commonly available IRAS image products: $4'$ for the
Infrared Sky Survey Atlas (ISSA, Wheelock et al.\ 1994), and $1'$ (after HIRES
reprocessing) for the IRAS Galaxy Atlas (IGA, Cao et al.\ 1997) and
Mid-Infrared Galaxy Atlas (MIGA, Kerton \& Martin 2000).  At these 
resolutions a collection of objects can appear as a single point
source.  Moreover, areas of extended dust shells surrounding
\ion{H}{2} regions can be misidentified as point sources.  While the
IRAS colors can be used as a general guide to the nature of the source
\citep[e.g.,][]{hug89} they can sometimes be misleading (see
Sections~\ref{sec:irasnorthwest} and \ref{sec:bstar}).

Moving from the infrared wavelengths observed by IRAS to the
submillimeter yields two advantages.  First, the current generation of
submillimeter mapping bolometers have far better resolution than the
IRAS data and sufficient sensitivity to allow star formation regions at
the distance of KR~140 to be investigated.  Second, the submillimeter
observations are sensitive to colder dust not detected by IRAS and thus
allow a more complete picture of the dust distribution.  The
observations we have obtained have both clarified the nature of the IRAS
point sources surrounding KR~140 and have led to the discovery of new
discrete sources not visible in the IRAS data.

In Section~\ref{sec:observations} the observations are described. Data
analysis techniques are outlined in Section~\ref{sec:analysis}.  In
Section~\ref{sec:discussion} the results of the analysis are discussed.
Finally, in Section~\ref{sec:conclusions} the conclusions are presented.

\section{OBSERVATIONS}
\label{sec:observations}

The observations were obtained through the CANSERV service observing
program at the James Clerk Maxwell Telescope (JCMT \footnote{The JCMT is
operated by the Joint Astronomy Centre on behalf of PPARC of the UK, the
Netherlands OSR, and NRC Canada.}) using the Submillimeter Common-User
Bolometer Array (SCUBA, Holland et al.\ 1999).  Observations using SCUBA in its
``jiggle map'' mode were obtained in December 1997 and observations in
the larger-scale ``scan mapping'' mode were obtained in July 1998.

SCUBA was used to acquire 450~$\mu$m and 850~$\mu$m data simultaneously.
At these wavelengths the beam can be approximated as a Gaussian with a
FWHM of about 8$\arcsec$ at 450~$\mu$m and 14.5$\arcsec$ at 850~$\mu$m
\citep{mat99}.  The main beam solid angle ($\Omega_{\rm MB}$) can then
be calculated using the standard formula $\Omega_{\rm MB} =
1.133\theta_{\rm FWHM}^{2}$ \citep{kra66}.

	\subsection{Jiggle Maps}
	\label{sec:jiggle}

Jiggle mapping, being closely related to pointed observations, allows
users to make images of objects on the scale of the SCUBA field of view,
$\sim2'$.  The northwest rim of KR~140 was selected as the target for
the initial jiggle maps because the brightest infrared source
(IRAS~02160+6057) is located there (see Section~\ref{sec:irasnorthwest}
for the analysis of this source). Figure~\ref{fig:jigmap} shows the
resulting jiggle maps ($\sim 2' \times 2'$ in size). A chop throw of $2'$ with a position angle of 135$\degr$ was used
to obtain a reasonable sky estimate by avoiding bright infrared emission
associated with the \ion{H}{2} region.  Beam sizes of 8.5$\arcsec$ and
14.5$\arcsec$ were achieved for this set of observations.  The average
normalized sky transparency during the observations was 0.81 at 850
$\mu$m and 0.34 at 450 $\mu$m.  The expected noise equivalent flux
density per pixel for the bolometer arrays is approximately 0.08 and
0.95~Jy~Hz$^{-0.5}$ at 850 and 450 $\mu$m respectively \citep{hol99}.
The measured noise levels in the maps are $\sim 0.01$ Jy/beam and $\sim
0.08$ Jy/beam at 850 and 450~$\mu$m respectively, consistent with the
expectations from sky transparency and bolometer characteristics.
There are no strong point sources visible in either map. Instead,
extended emission is quite evident in the 850 $\mu$m map (S/N $\sim$
4--8).  The same region of emission is barely visible in the 450 $\mu$m
map.  However, convolving the image to the 850 $\mu$m resolution reduces
the noise level to 0.05 Jy/(850 $\mu$m beam) and results in the extended
emission being clearly visible (S/N $\sim$ 3--7).

	\subsection{Scan Maps}
	\label{sec:scan}

Larger scale maps of KR~140 at 450/850~$\mu$m were obtained in July 1998
using the recently (at the time) developed ``scan mapping'' observing
mode which allows larger areas to be mapped out by scanning the array
across the sky.  These maps included the jiggle-map region and enabled
the study of the other IRAS point sources near KR~140.

The raw bolometer data were reduced using the standard SCUBA software
\citep{hol99} for flatfield and extinction corrections.  The data were
then further reduced and transformed into the final maps using a matrix
inversion technique \citep{joh00a} which effectively minimizes the
difference between the individual chop measurements and the resulting
map.  One advantage of this technique compared to standard reductions is
that the bolometer input may be weighted to account for variations in
the quality of the measurement.  Figure~\ref{fig:scanmaps} shows maps created at this stage.  Weiner filtering these maps \citep[see,
e.g.,][]{pre92} removes the high (spatial) frequency noise associated
with the individual chop measurements without affecting structure on
scales equal to or larger than the JCMT beam.  Further processing of the
map proves useful.  Due to the finite chop size, SCUBA maps have their
low spatial frequency modes suppressed.  Restoring these modes requires
a large re-amplification which also amplifies any existing low frequency
noise.  Furthermore, correlated errors introduced by variations in the
sky opacity and the drift in the bolometer response produce excess
residual low frequency modes.  Thus the large scale structure in the map
is less reliable, and in order to better visualize the small scale
features in the data, the low frequency modes were entirely suppressed.
This was achieved by convolving the filtered map with a $\sigma =
130\arcsec$ Gaussian beam (twice the largest chop throw) and then
subtracting the convolved map from the original to remove any
large-scale structure.  Both images were then brought to a common
20$\arcsec$ resolution to facilitate comparison.  In
Figure~\ref{fig:scanmapsmooth} the resultant maps are displayed.
Noise levels are $\sim$ 0.01 Jy/beam and 0.05 Jy/beam at 850 and 450~$\mu$m
respectively.

Figure~\ref{fig:scancomp} compares the derived 850~$\mu$m scan map for the
region with other data obtained from the CGPS database (see
\citet{bal00} for details).  

\section{ANALYSIS}
\label{sec:analysis}

Our quantitative analysis focuses on the discrete sources detected
in our maps.  The next three subsections describe how the sources were
detected, and the subsequent measurements of flux, size, temperature,
mass and density for these objects.  Interpretation of these
results is presented in Section~\ref{sec:discussion}.
	
	\subsection{Source Detection, Flux and Size}
	\label{sec:detect_flux_size}

Individual sources were identified through a combination of visual
inspection of the maps, and by using a 2D version of the automated clump
finding algorithm CLUMPFIND \citep{wil94} which separates objects along
minimum flux surfaces within the map.  In total 22 sources were
detected; 11 at both 850 and 450~$\mu$m, nine sources at only
850~$\mu$m, and two at only 450~$\mu$m.  These last two sources may be
visible at 850~$\mu$m but it is hard to be certain because of the
increased noise at that location due to edge effects from processing.
The flux-weighted centroid of each source was determined and used as a
measure of the source position.  We have assigned the acronym KMJB to
these new submillimeter sources. Except for one of our sources
(KMJB~1 = IRAS 02174+6052) detected near the edge of the map, there was no
significant difference in the centroid positions between the two
observed wavelengths (when applicable).  For KMJB 1 we have reported
the 850~$\mu$m centroid position because of the higher quality of the
data.  Table~\ref{tbl:sources} lists the positions of the sources
indicated by the white crosses in Figure~\ref{fig:scanmapsmooth}.

In order to estimate the size of the sources, an azimuthally averaged
Gaussian profile was fitted to each source using the image analysis
program kview \citep{goo95}.  To take into account the beam size, a
simple convolution of a Gaussian source with a Gaussian beam was
assumed, $\theta_{\rm observed}^2 = \theta_{beam}^2 + \theta_{\rm
object}^2$, where $\theta_{beam} = 20\arcsec$.  Physical sizes ($D_{\rm
eff} = \theta_{\rm object} d$) were calculated assuming a distance $d=
2.3\pm$0.3 kpc to KR~140 and tabulated in Table~\ref{tbl:sources}.  The
effective diameters range from 0.7 pc down to the resolution limit,
0.1 -- 0.2 pc. This size scale of a few $\times$ 0.1 pc is typical for
cores observed within molecular clouds \citep{eva99}.

Flux measurements of the sources were obtained using the Imview software
package\footnote{Imview is a multi-purpose image display and analysis
package developed at DRAO.}.  The uncertainty in each measurement is dominated
by the uncertainty in choosing a correct background level. To quantify this uncertainty, we interactively selected three
different bounding polygons for each source, and used two different
functional forms for the background (twisted plane and twisted
quadratic) which were fit to the boundary points of the polygon.  The
six resulting flux measurements were then averaged together and the
standard deviation of the measurements was calculated and used as a
conservative uncertainty estimate.  The results are summarized in
Table~\ref{tbl:sources}.

	\subsection{Temperature}
	\label{sec:temp}

The spectral energy distribution (SED) of the dust emission was
parameterized using what we call a ``bluebody'' spectrum after
Dent, Matthews, \& Ward-Thompson (1998): a single temperature Planck function 
with a $\nu^2$ dependence ($\beta = 2$) for the dust emissivity
\footnote{``Modified  blackbody'' is the term often used in the
literature, but bluebody is both shorter and more descriptive.}.  While this 
is clearly a gross oversimplification of the dust properties \citet{den98} and
\citet{gor95} point out that the quantities derived from these simple
fits are often consistent with results derived from a more complex, and
often non-unique, analysis of the data.  

For the isolated sources detected by IRAS, this SED was fit to the
submillimeter and 100 and 60 $\mu$m IRAS data.  However, for two
sources, IRAS~02168+6052 and IRAS~02157+6053, this technique was not
possible because of the significant beam size difference between IRAS
and SCUBA: multiple submillimeter sources were detected within the
larger IRAS beam.  In these cases, and for sources without IRAS
counterparts, the submillimeter photometry was used where possible to
derive a 450/850~$\mu$m bluebody color temperature.  This estimate can
be rough for two reasons.  For warmer dust, one is out on the
Rayleigh-Jeans tail of the Planck function which is relatively
insensitive to temperature. The calculated
temperature for colder dust is still very uncertain due to the
relatively poor quality of the 450~$\mu$m data.

For sources detected at only a single wavelength representative masses
were calculated using a temperature of 20 K, as has been done in other
single wavelength analyses of millimeter and submillimeter data (e.g.,
Johnstone et al.\ 2000b, Motte, Andr\'{e}, \& Neri 1998). This
temperature is a typical value observed for compact dust features
found by submillimeter studies of the Orion molecular clouds \citep{chi97, lis98}.

	\subsection{Mass and Density}
	\label{sec:mass}

Table~\ref{tbl:sources} summarizes the mass and density estimates.
 The mass calculation used the standard
formulation of \citet{hil83} for an optically thin molecular cloud with
a uniform temperature:
\begin{equation}
\label{eqn:scuba_mass}
M = d^2F_\nu C_\nu / B_\nu(T)
\end{equation}
where $M$ is the total (dust and gas) cloud mass, $d$ is the distance to
the objects, $F_\nu$ is the observed flux density, $B_\nu(T)$ is the
Planck function evaluated at dust temperature $T$, and $C_\nu$ is a
factor combining the dust opacity ($\kappa_\nu$) and the dust to gas
mass ratio ($C_\nu= M_g/(M_d \kappa_\nu)$).

Table~\ref{tbl:cvals} summarizes various values of $C_{850}$ derived
from the literature.  Most of the articles cited report the value of
$C_\nu$ at 1~mm, and the values shown have been calculated assuming that
$C_\nu \propto \nu^{-2}$ ($\beta = 2$ for the opacity).  The canonical
\citet{hil83} value is derived from the observations of the reflection
nebula NGC 7023 (Hildebrand 1983).  While this is empirically derived,
ideally it should be applied to observed objects which have similar
properties to the calibration object \citep{hil83}; this is probably not
the case here where we find lower temperatures and higher densities.
The \citet{dra84} value is appropriate for dust in the general ISM;
however, it is known that properties of dust in dense clouds differ from
those in the diffuse ISM (Kim, Martin, \& Hendry 1994).  The difference in optical
properties arises from a combination of coagulation, leading to an
increase in grain porosity, and the accretion of volatile grain mantles.
The remaining values in the table are from various models that have
attempted to deal directly with these processes.  Various authors
suggest different values ranging at 850~$\mu$m from 21 -- 86 g
cm$^{-2}$, depending upon the gas density and temperature.  Since we do
not know the density in advance we have adopted a value of $C_\nu=50$ g
cm$^{-2}$ at 850$~\mu$m and, assuming a $\beta = 2$ dust emissivity,
$C_\nu=14$ g cm$^{-2}$ at 450$~\mu$m for this investigation.  This is in
general agreement with the results of a recent detailed observational
study of circumstellar material by \citet{vdt99}.

The masses derived in our study range from 0.7 to 130~M$_\odot$.  They
can be linearly scaled to another value of $C_\nu$ if desired.  The
uncertainty in the distance to KR~140 leads to a systematic 26\%
uncertainty in the derived masses.  The average tabulated uncertainty in the
flux density measurements is 17\% at 850~$\mu$m and 28\% at 450~$\mu$m.
Uncertainties relating to the derived temperature are harder to quantify
because of the simple bluebody model that we have used and because the
temperature enters into the mass derivation non-linearly through the
Planck function.  Figure~\ref{fig:terr} shows the change in the mass
derived from 850~$\mu$m flux measurements produced by varying the
temperature.  At lower temperatures the derived mass is very
sensitive to the temperature.  For example, reducing the temperature by
only 2~K will cause the derived mass to almost double.  The same
absolute change at 15~K causes an increase of only 28\% and at 25~K the
change is only 12\%.  Based upon these considerations a total
uncertainty (not including the uncertainties in $C_\nu$) for the derived
masses of $\pm 50$\% is an optimistic minimum value \citep[see,
e.g.,][]{gor95}.

For each source the cube of $D_{\rm eff}$ was taken as a rough estimate
of the volume and used to calculate a density.  Considering both the
lack of information about the true three-dimensional structure of the
sources and the uncertainty in the masses, these densities should be
considered as only order of magnitude estimates.

\section{DISCUSSION}
\label{sec:discussion}

In this section we first consider what the new submillimeter
observations can tell us about the nature of the IRAS point sources,
drawing on the multifrequency data available.
Discussion about the newly detected sources and the region as a whole
follow in Sections~\ref{sec:massfunc}, \ref{sec:bonnor} and
\ref{sec:trigstar}.

	\subsection{The Nature of the IRAS Sources}
	\label{sec:iras}

	\subsubsection{IRAS 02160+6057 = KMJB 15}
	\label{sec:irasnorthwest}

The source IRAS~02160+6057 identified with the north-west dust arc has
been the subject of two molecular line investigations.  \citet{wou89}
identified it as a potential star-forming area via its IRAS colors (see
their paper for the exact selection criteria) and examined it (along
with about 1300 other IRAS point sources) for CO emission.  They found a
CO feature in that direction at a velocity of $-$49.7~km~s$^{-1}$ (LSR),
which corresponds to CO in the associated background molecular cloud
\citep{bal00}, and assigned it the catalog number [WB89]417.  This point
source was not detected in a subsequent H$_2$O line survey by Wouterloot, 
Brand, \& Fiegle (1993).

A comparison of the scan maps with the IRAS images of KR~140 for this
region provides a striking example of how much the morphology of the
dust emission associated with an \ion{H}{2} region can change with
wavelength.  In the IRAS images this region at all bands
(Figure~\ref{fig:scancomp};
Ballantyne et al.\ 2000, Figure~2)
is the dominant feature.  In
contrast, while the feature is visible in the submillimeter maps it is
certainly not the dominant one.  This contrast indicates that the dust
emission one sees from this area is from relatively hot dust with no
additional cool components along the line of sight.  This is consistent
with the lack of significant clumped CO emission in this area which
could have indicated a region containing colder dust grains (see
Figure~\ref{fig:scancomp}).  The submillimeter and infrared emission are
also aligned spatially which is what we would expect if the
submillimeter emission is just coming from the hot dust we see in the
IRAS bands (see Figure~\ref{fig:jigmap}).  Morphologically,
IRAS~02160+6057 appears to be part of the partial dust shell surrounding
KR~140.
This line of sight has one of the highest column densities of {\it warm} dust
in the KR~140 \ion{H}{2} region and so it is possible that a protostar
could be forming there as a result of the expansion of the \ion{H}{2}
region.  However, examination of the HIRES images shows no point-like
features in the dust arc and none are found in the new submillimeter
observations.

The average mean flux density at 850~$\mu$m is $0.7\pm0.1$ Jy in the
jiggle map and $1.0\pm0.3$ Jy in the scan map.  For our calculations we
have used 0.8$\pm$0.2~Jy as the flux density at 850~$\mu$m. At
450~$\mu$m we used the flux density from the jiggle maps, $4\pm1$
Jy, since the feature is not seen in the 450~$\mu$m scan maps. 
In order to construct an SED for the source we measured infrared flux
densities in the same region at 60 and 100~$\mu$m using HIRES images of
KR~140.  We found F$_{60} = 42 \pm 5$~Jy and F$_{100} = 101 \pm 8$~Jy.
Combining this infrared data with the submillimeter data we obtain a
best-fit bluebody curve to the SED with T$=27.5$~K (see
Figure~\ref{fig:sourcesed}).  Comparison with the azimuthally averaged
dust temperatures for KR~140 (see Figure~5 in Ballantyne et al.\
2000), shows that our derived temperature is consistent with dust
being heated by the central star VES~735.

We conclude that IRAS~02160+6057 is not a point source but simply part
of the dust shell associated with KR~140.  This naturally explains the
non-detection of H$_2$O in the \citet{wou93} study.  Using equation 1 we
find a mass of 9 M$_\odot$ in this segment of the shell; the density
there is 10$^{2.9}$ cm$^{-3}$, the lowest for all of the sources and would
be even lower if allowance were made for limb brightening (line of sight
distance $>D_{\rm eff}$).  The
structure of this dust shell (morphology, density, temperature) was
examined in detail in Section~6 of \citet{bal00}.

	\subsubsection{IRAS 02168+6052 =  KMJB 6, 7 and 8} 
	\label{sec:iraseastrim}

At submillimeter wavelengths the region around the single IRAS point
source IRAS 02168+6052 contains three sources (see
Figure~\ref{fig:eastshell}).  The brightest submillimeter source in the
area (KMJB 8) is well off the 60~$\mu$m centroid and is associated with
the CO emission.   It is also detectable at 450~$\mu$m.  The resulting
color temperature is 8$\pm$2 K. This leads to a mass of 80 M$_\odot$
and a density of 10$^{4.1}$~cm$^{-3}$.

KMJB 6 breaks up into three discrete objects at 450~$\mu$m.  We
calculated a 450/850 color temperature using the combined flux of the
sources at 450 and the measured flux density at 850 $\mu$m.  The derived
dust temperature is 7$\pm$2 K.  The corresponding mass is about 55
M$_\odot$, and the overall density in this region is around
10$^{4.3}$~cm$^{-3}$.  KMJB 6 and 8 are most likely cold molecular
cores; however, the very cold temperatures derived for these objects
should be confirmed with future spectral line observations.

KMJB 7 is located near the position of IRAS~02168+6052 and is most
likely a dense knot of dust and gas within the dust shell surrounding
KR~140.  Like in KMJB 15, the shell is the source of the prominent
ridge of emission seen in all of the IRAS emission bands.  Faint
submillimeter emission from the dust shell is visible as a diffuse ridge
running between KMJB 6 and 8.  The temperature of dust within this
shell was measured to be around 30~K \citep{bal00}, which is consistent
with the ratio of 60 and 100~$\mu$m to 850~$\mu$m fluxes.  Using this
dust temperature we derive a mass of 1 M$_\odot$ for the knot that is
KMJB 7.

	\subsubsection{IRAS 02157+6053 = KMJB 17, 18 and 19}
	\label{sec:molcore}

This region (see Figure~\ref{fig:region5}) is associated with
molecular material located behind KR~140 \citep{bal00}. The three
submillimeter sources are located within the larger CO envelope running
roughly SW--NE, following the orientation of the 60~$\mu$m contours.
The IRAS scan pattern for this region places the major axis of the
detectors along a SW--NE direction and so the PSC fluxes are the
contribution of the ensemble of sources in the area.  The crowded nature
of the region makes it difficult to use the IRAS flux densities to
determine temperatures.

Nevertheless, KMJB 19 is well off the $60~\mu$m centroid and hence
relatively cold.
It is  detectable at 450~$\mu$m, leading to a 450/850
$\mu$m color temperature of 8$\pm$2~K.  The resulting mass (50
M$_\odot$) and density 
(10$^{4.6}$ cm$^{-3}$) are consistent with the
object being a cold molecular core, like KMJB 8.

Mass estimates were obtained for KMJB 17 and 18 using the 850~$\mu$m
flux and assuming a temperature of 20~K (which is roughly consistent
with explaining the $60~\mu$m emission).  The mass and lower limits on
the density (the 
sources are point sources) are both consistent with the
objects being molecular cores.  KMJB 18 is probably warmer than KMJB
17 given the facts that they are of comparable brightness at 850~$\mu$m
and 18 is located closer to the centroid of the 60~$\mu$m emission.
Taking this into account would accentuate the mass and density differences.

	\subsubsection{IRAS 02171+6058 = KMJB 3} 
        \label{sec:bstar}

This source is located to the northeast and outside of the KR~140 radio
continuum emission.  It seems likely that this relatively bright source
is associated with molecular material at the distance of KR~140.  On the
basis of the SED, this would be a class I object \citep{and93}.  We
proceed with the analysis assuming a distance of 2.3 kpc and show that
this consistently explains a number of observational characteristics of
the source.
 
The infrared colors of the infrared point source IRAS~02171+6058 are
characteristic of an ultracompact \ion{H}{2} (UCHII) region based upon
the criteria of \citet{woo89}: $\log(F_\nu(60)/F_\nu(12)) > 1.3$ and
$\log(F_\nu(25)/F_\nu(12)) > 0.57$.  They also match the criteria of
\citet{hug89} for an \ion{H}{2} region: $\log(F_\nu(60)/F_\nu(25)) >
0.23$ and $\log(F_\nu(25)/F_\nu(12)) > 0.40$.  It should be noted that
the PSC 60~$\mu$m flux is only of moderate quality (as defined in the
IRAS Explanatory Supplement 1988), but it is in agreement with a flux density measurement
we made from the HIRES images of the region.  The source does not meet
the additional flux criteria set by \citet{hug89}, F$_{\rm 100} \geq 80$
Jy which was designed to help filter out reflection nebulae and
extragalactic objects.  The source was included in a CS (2-1) survey by
Bronfman, Nyman, \& May (1996) and a methanol maser survey by
\citet{lyd97}, but in both cases no detections were made.

Since this IRAS point source is detected at 12, 25, 60, 100 and
850~$\mu$m it is surprising that it is not seen at 450~$\mu$m.  We fit a
single-temperature bluebody to the 850, 100 and 60 $\mu$m data points to
obtain a temperature estimate of 27~K (see Figure~\ref{fig:sourcesed}).
This best fit predicts a flux density of 2.8 Jy at 450~$\mu$m which
normally should be detectable, especially in the smoothed scan map.
However, there is a large gradient in the 450~$\mu$m scan map around
this position due to enhanced noise at the edge of the map which
may be making it hard to detect the source.
The fitted SED yields a total mass of 5 M$_\odot$.  Using 0.23~pc as a
typical length scale, the density is found to be
$10^{4.1}$ cm$^{-3}$.

Note that this dust is not heated by the exciting star of KR~140; there
must be a source internal to the cloud.  The total flux derived from the
best-fit SED gives a luminosity of $L=10^{2.5}$ L$_{\odot}$ assuming a
distance of 2.3~kpc.  This can be compared to the bolometric luminosity
for main sequence B5 V -- B8 V stars which
ranges from $L\sim10^{3.0}$ to 10$^{2.3}$ L$_{\odot}$ \citep{lan92}.

This is consistent with the lack of a
corresponding point source in the 1420 MHz continuum image.  The
sensitivity of the 1420 MHz CGPS images is 0.2 mJy/$1'$ beam
\citep{tay99}.  Ionization equilibrium can be used to relate this
brightness level to the rate at which a star is emitting ionizing
photons (Q).  At the distance of KR~140 this brightness level
corresponds to log(Q) = 44.0 for a point source, and so any embedded O or
early B main sequence stars (which have log(Q) $\geq$ 47; Kerton 2000) would
certainly be detected in the CGPS radio data.

Chini, Kr\"{u}gel, \& Wargau (1987) derived a relationship between the total 
luminosity (as seen reradiated by dust) of a compact \ion{H}{2} region and 
the gas mass:
\begin{equation}
\label{eqn:scuba_chini}
L = (56\pm21)M_g^{0.93\pm0.06}
\end{equation}
where $L$ and $M_g$ are the luminosity and gas mass in solar units.  The
lowest luminosity and mass considered in their study was $L= 10^{3.6}$
$L_{\odot}$.  Using an extrapolation of this relationship, the
luminosity derived for this object above corresponds to a mass of
6$^{+6}_{-2}$ $M_{\odot}$.  This is consistent with the independently
determined measurement of the mass based upon the submillimeter
observations, and indicates that the \citet{chi87} relationship may hold
to lower $L$ and $M$ values than covered in their study of regions with
ionizing stars.

	\subsubsection{IRAS 02174+6052 = KMJB 1}
	\label{sec:h2new}

The IRAS PSC fluxes for this object are only of marginal use since they
contain two upper limits at 60 and 100~$\mu$m.  To get more information
we examined HIRES images of the KR~140 region.  At 100~$\mu$m the source
is not visible against the very strong emission coming from the nearby
dust shell of the \ion{H}{2} region.  We were able to measure flux
densities at the other three IRAS bands.  At 60~$\mu$m we obtained a
flux density of 12$\pm$2 Jy.  This is consistent with the upper limit
reported in the PSC.  Flux densities obtained at 12 and 25~$\mu$m agreed
with the PSC values.  Like KMJB 3, this has a class~I SED.
The object does have $F_\nu(60)/F_\nu(25)$ and $F_\nu(25)/F_\nu(12)$
colors in the range expected for an UCHII region as set by
\citet{hug89}, although the $F_\nu(25)/F_\nu(12)$ color is at the lower
limit.  \citet{woo89} used the $F_\nu(60)/F_\nu(12)$ color and a
higher limit to the $F_\nu(25)/F_\nu(12)$ color to define UCHII
regions, and this object does not match their criteria for an UCHII
region.  Therefore, from the IRAS colors alone it is unlikely that the
object is an UCHII region. 

We fit a single-temperature bluebody to the 850, 450 and 60 $\mu$m data
points to obtain a temperature estimate of 22~K (see
Figure~\ref{fig:sourcesed}).  The best-fit curve predicts a flux density
of 68~Jy at 100~$\mu$m which is consistent with the upper limits
reported in the PSC.  This value is also below the 100~$\mu$m flux
density limit set by \citet{hug89}.  Values from this best-fitting curve
were then used to derive a total mass of 19 $M_\odot$.

Using the SED, and assuming a distance of 2.3~kpc, we calculated a total
luminosity of 10$^{2.8}$ $L_\odot$.  As with KMJB 3 there is no
corresponding 1420 MHz continuum source.  This is what one would expect
for a star with luminosity corresponding to B5 V.  Using this luminosity
and the mass-luminosity relation of \citet{chi87}
(equation~\ref{eqn:scuba_chini}) we obtain a mass estimate of
14$^{+14}_{-5}$ $M_\odot$ in agreement with the value derived from the
submillimeter observations.  

	\subsection{Cumulative Clump Mass Function}
	\label{sec:massfunc}

The cumulative mass distribution for the sources detected in this study
is shown in Figure~\ref{fig:clumps}.  A least-squares fit to the data of
a $N(>M) = N_o M^{-\alpha}$ power law resulted in $\alpha= 0.49\pm0.04$.
\citet{kra98} compiled results from a number of studies and showed that
a power law distribution with $\alpha = 0.65\pm0.10$ describes the clump
mass distribution in molecular clouds for clumps with masses ranging
from $10^{-4}$ to several $10^{4}$~$M_\odot$. The two studies contained
within \citet{kra98} that are closest to our study with respect to the
mass range covered agree within the stated uncertainties.  Clumps in the mass
range 0.6 to 160~$M_\odot$ surrounding the \ion{H}{2} region S~140 were
found to follow an $\alpha= 0.65\pm0.18$ power law, while clumps in the
mass range 0.8 to 50~$M_\odot$ surrounding the \ion{H}{2} region NGC
1499 followed $\alpha=0.59\pm0.18$.  In Figure~\ref{fig:motte_clump} we
have plotted the cumulative mass distribution for the larger scale
cores (0.05 pc to 0.3 pc in size) in $\rho$ Oph using the data from
\citet{mot98}. These objects also follow a power law relationship
similar to what we find in KR~140, $\alpha= 0.6 \pm 0.08$.
 
In contrast, millimeter and submillimeter studies of the nearby
$\rho$~Oph region have shown that the mass distribution of the smaller
scale fragments (2400 to 5000 AU in size) contained within the cores
appears to mimic the stellar IMF \citep{mot98,joh00b}. \citet{tes98}
used the OVRO interferometer to study similarly sized objects in the
more distant Serpens molecular core.  They also found a steep power law 
distribution consistent with the stellar IMF, $\alpha = 1.1$.  The key similarities in these three studies are that they probed sufficiently small linear
scales (0.01 -- 0.02 pc) and that a sufficient range in masses was
observed to detect a steepening of the mass function above 0.5 $M_\odot$.
These both appear to be important requirements.  Of the studies listed
by \citet{kra98} only the study of the nearby ($d = 65$~pc) quiescent cloud
L1457 has sufficiently high angular resolution to be comparable to the
$\rho$ Oph and Serpens studies.  However, the highest mass clump
detected is only 0.2 $M_\odot$ and so all of the detected objects lie on
a shallow part of the mass distribution function ($\alpha =
0.77\pm0.3$). One can interpret this as either being consistent with
the low mass portion of the stellar IMF or being consistent with a
global power law for molecular cloud structure. Our study of KR~140 has
sufficient overlap in mass coverage to probe the connection between
cloud structure and the stellar IMF, but the larger
scale objects that we detect at our relatively low resolution follow the
well-established $\alpha\sim0.5$ power law seen for structures
within giant molecular clouds and for the clouds themselves (Williams,
Blitz, \& McKee 2000).

	\subsection{Nature of the Sources}
	\label{sec:bonnor}

How might the submillimeter sources relate to future star formation?  
The Bonnor-Ebert (BE) critical mass \citep{ebe55,bon56} is given by
\begin{equation}
\label{eqn:bemass}
M_{BE} = \frac{2.4 R a_{s}^{2}}{G}.
\end{equation}
Using $R = D_{\rm eff}/2$ and the standard formula for the isothermal
sound speed ($a_{s}$) we can write
\begin{equation}
\label{eqn:bemass2}
M_{BE} = 0.99 \left(\frac{D_{\rm eff}}{\rm pc}\right)
\left(\frac{T_{gas}}{\rm K}\right) .   
\end{equation}
To derive $M_{BE}$ from our data we have assumed that $T_{gas} =
T_{dust}$ since our CO data are of too low resolution to give
information about the gas dynamics or temperatures within specific
clumps directly.

Figure~\ref{fig:massbe} plots the derived submillimeter mass $M$ versus
the corresponding $M_{BE}$ for each clump. Clumps with $M < M_{BE}$ are
not gravitationally bound or unstable.  Prominent among these are the
sources that we have identified as parts of the dust shell, 7 and 15;
their existence depends on pressure of the \ion{H}{2} region and its
dynamical expansion.  All of the other ``stable'' sources (2, 10, 13,
16, 20, 21, 22) would also be pressure supported or transient.  There are
clumps which are detected at only one submillimeter wavelength and have
no clear IRAS counterparts.  For these we have assumed $T =
20$~K. Adopting a lower $T$ would have led to a lower $M_{BE}$ and
higher $M$ (Figure~\ref{fig:terr}), moving the clumps toward the region of
instability.  However, unless we adopted a substantially lower
temperature ($\sim$ 7 K) the clumps would remain in the stable region.

Clumps with $M > M_{BE}$ are massive enough to collapse under the effect
of self-gravity.  Excluding the clumps that have already formed stars
(KMJB 1 and 3), we find eleven, including the aligned sources 11, 12,
and 14, and 4, 5, 6, 8, 9, and 19.  This result derives from the low
temperature ($T < 13$~K) that we have measured for these sources.
KMJB 17 and 18 are also barely over the instability line.  For these
we have assumed $T = 20$~K.  Adopting a lower $T$ would move the clumps
further into the region of instability. 

By combining the available observational data on molecular clouds, \citet{lar81} showed that over a wide range in sizes
gravitationally-bound molecular clumps follow the relationships $\sigma
\propto D^{0.4}$ and $\sigma \propto M^{0.2}$ were $\sigma$ is the
velocity dispersion, $D$ is the largest linear dimension (in pc), and
$M$ is the mass of the object; thus $M \propto D^{2}$.  This is
equivalent to saying that the objects have a constant column density.
Figure~\ref{fig:massrad} shows $M/D^{2}$ (or equivalently
$A_V$) vs.\ $D$.  The bimodal distribution of the sources is
apparent. All of the unbound sources, with $M < M_{BE}$, have a very
low column density below $A_V \sim 2$. The average column density, $nD
= 2.1 \times 10^{22}$~cm$^{-2}$, for all of the sources with $M >
M_{BE}$ (except the outlying sources 5 and 12) is twice
what would be deduced from \citet{lar81}. This may be due to our using
a different tracer, and thus a different length and mass scale.
\citet{mye83} showed that this result can be interpreted as virial 
equilibrium with non-thermal pressure support from turbulent gas motions.

As discussed by \citet{eva99}, star formation appears to  occur only
in cores with $N_{\rm H} > 8\times10^{21}$ cm$^{-2}$ ($A_V \sim 4$).  
Theoretically this is expected because of the sharp decrease in the ionization
fraction within such cores  which in turn allows
ambipolar diffusion to proceed at a faster rate \citep{mck89,mck99}.
In clouds where both magnetic and turbulent support are important one
still needs a low ionization fraction to allow collapse to proceed
after the loss of turbulent support \citep{shu99}.  The sharp decrease in
the ionization fraction at $A_V \sim 4$ may also be related directly
to the loss of turbulent magnetohydrodynamic support because of the
reduction in ion-neutral coupling \citep{ruf98}.
\citet{oni98} studied 40 molecular cloud cores in Taurus in C$^{18}$O  and
found  that all of the cores with $N_{\rm H} > 8\times10^{21}$
cm$^{-2}$ were associated with H$^{13}$CO$^{+}$ emission
(which they interpret as being protostellar condensations) and/or
cold IRAS sources.  In our sample of sources around KR~140 only
KMJB 1 and 3 have unambiguous evidence for star formation having
occurred and both are high column density sources.  It would be interesting to
investigate the other observed sources for evidence of star formation
via higher resolution observations of dense gas tracers. KMJB 12,
lying well above the line, is an interesting object for
further study because of its large mass and location on the ridge
boundary discussed below.  Among the clumps with $M > M_{\rm BE}$, KMJB 5 has
an unusually low column density.

	\subsection{Clump and Star Formation in the KR~140 Region}
	\label{sec:trigstar}

Are the clumps just structure in the cloud that pre-existed the
formation of VES~735 and KR~140 or is there ongoing evolution?  The
three high density sources 11, 12, and 14 are of particular
interest.  Their cold temperature, small size, and high density
suggest that they are
cold molecular cores.  They are aligned along the edge of a CO cloud
with V$_{\rm LSR} = -46.0$ km s$^{-1}$ (Figure~\ref{fig:scancomp}) and also
parallel to, but outside 
of, a ridge which appears in a high resolution H$\alpha$ map
\citep{ker99}.
These sources are probably not foreground to the  \ion{H}{2} region
since even with their large column densities they do not produce
distinctive silhouettes on the H$\alpha$ map.
This and the striking alignment suggest that a causal relationship
exists between the formation of these clumps at this location and the
expansion of the KR~140 \ion{H}{2} region.
We have estimated that the age of the KR~140 \ion{H}{2} region is about
$2 \times10^{6}$ years \citep{bal00}, which is certainly sufficient time
for a layer of gas to be swept up and become gravitationally unstable
even for the low density molecular cloud associated with KR~140.
Other cold submillimeter sources (e.g., 6, 8, and 19) lie at the the
edge of the \ion{H}{2} region (see Figure~\ref{fig:scancomp}) and their
location is also suggestive of a link between the expansion of the
\ion{H}{2} region and their formation.  

This mix of clumps, some that appear to have formed in particular
locations because of the action of other stars, some that appear to have
formed spontaneously in various regions throughout a larger cloud, and
some which are forming stars (this would include the clump that formed
VES~735), is not unlike what is seen in other star-forming regions
(e.g., for $\rho$~Oph, see the discussion in Section 6.3 of Motte et
al.\ 1998).

It would be expected that a cluster would be forming near VES~735 and so
it is interesting to examine the stellar (or proto-stellar) content.
Our analysis has shown that there are no UCHII regions in the vicinity
of KR~140 and so VES 735 appears to be the only O star to have formed in
the region.  From \citet{bal00} the total mass of the molecular cloud 
associated with KR~140 is estimated to be about
5$\times10^{3}$~$M_\odot$.  O star formation in clouds with a
cloud mass of below 10$^4$ $M_\odot$ is considered rare \citep{elm85,
wil97} and so it would actually have been very unusual to find yet
another O star in the vicinity of VES 735.

KMJB 1 and 3 contain embedded objects with the luminosity of main
sequence B stars.  Their position in the molecular
cloud suggests no causal relationship with the \ion{H}{2} region and
instead they could have formed in the original molecular cloud along
with VES 735.  The evolution time for a 4--6 $M_\odot$ star ($\sim$ B5 V
to B8 V) to the main sequence ranges from 0.4 -- 8.4 $\times10^{5}$
years \citep{pal93}.  \citet{bal00} estimate that VES 735 and KR~140 are
1--2 $\times10^{6}$ years old, ample time for these stars to have formed and
evolved to their present state; in fact their evolution is probably
relatively delayed rather than coeval.

Close to KR~140 there are five bright infrared stars (BIRS 128 to 132;
Elmegreen 1980; see Figure~14 in Ballantyne et al.\ 2000) with very red 
$m_R - m_I$
colors (4.3 to 6.7; $m$ and $M$ in this paragraph refer to apparent
and absolute magnitudes respectively). 
BIRS 128 is just outside of KMJB 7 (see Figure~\ref{fig:eastshell}), BIRS 131 and 132 are near KMJB
16, and BIRS 129 and 130 are seen projected against the \ion{H}{2}
region, north of VES~735. However, while \citet{elm80} considers
the stars as possibly ``embedded'' there is no significant submillimeter
(or infrared) emission coincident with their positions, which puts a
limit $A_V \sim 1$ or $E_{R-I} \sim 0.27$ on associated clumped dust.
The general foreground extinction seen to affect VES~735 and KR~140
amounts to $A_V \sim 6$ \citep{ker99} or $E_{R-I} \sim 1.6$.  They are
possibly more highly extinguished by the molecular material that lies
behind KR~140 ($V = -46$ to $-54$~km~s$^{-1}$).  Using the CO and H~I
data cubes and gas to dust conversion factors from \citet{dig96} or
\citet{str96} $E_{R-I} \sim 3$ seems possible.
Nevertheless, the stars probably have intrinsically redder colors than
early main sequence stars ($M_R-M_I \sim 0$).  They are not ionizing
stars.  They could be pre-main sequence stars or evolved giants, though
the latter seems less likely since these stars would have had lower mass
than VES~735 yet have evolved sooner.
Additional photometry at J, H, and K, when available from 2MASS, should
help to clarify both this and the overall stellar content of the recently
formed cluster.

\section{CONCLUSIONS}
\label{sec:conclusions}

Based on the analysis of the SCUBA 450 and 850~$\mu$m maps of the KR~140
region we conclude that:

1) The putative UCHII region IRAS~02160+6057 is actually part
of the partial dust shell surrounding KR~140, to the NW.
IRAS~02168+6052 is also part of the partial dust
shell, to the east.

2) The infrared point source IRAS~02157+6053 appears in the
submillimeter as a number of discrete sources or clumps, some colder
than the IRAS source.

3) The cumulative mass distribution of clumps follows a $N(>M) =
N_oM^{-\alpha}$ curve with $\alpha= 0.49\pm0.04$.  This is consistent
with studies of clumps of this scale in other molecular clouds.  At the
distance of KR~140 we lack sufficient resolution to study the mass
distribution of even smaller cores, such as has been done for the more
nearby star forming regions of $\rho$ Oph \citep{mot98,joh00b},
 and Serpens \citep{tes98}.  The use of current (OVRO)
and future (SMA, ALMA) mm arrays is needed to build up a significant
sample of star forming regions and to probe different star forming
environments at sufficient resolution to probe the possible relationship
between observed structures in molecular clouds and the observed stellar
IMF.

4) The evolution of some of the clumps appears to have been affected by
the KR~140 \ion{H}{2} region.  Three molecular cloud cores are aligned
along the edge of a larger CO structure at the interface between the
molecular and ionized gas.  The alignment strongly suggests a causal
relationship to the expansion of the \ion{H}{2} region.

5) Many of the the submillimeter sources we detect are gravitationally
bound.  Most of the sources with $M > M_{\rm BE}$ follow a mass--size
relationship expected for objects in virial equilibrium with non-thermal
pressure support.  Upon the loss of non-thermal support they could be
sites of star formation.

6) A cluster of stars is probably forming along with VES~735.
IRAS~02171+6058, previously thought to be an UCHII region on the basis
of its PSC colors, has the luminosity of a lower mass embedded B star
(later than B5 V).  IRAS~02174+6052 is also an embedded star with
similar luminosity ($\sim$ B5 V).  Both objects appear to have formed
spontaneously in the molecular material outside KR~140.  Their evolution
is somewhat delayed relative to the evolution of the O star VES~735 and
its \ion{H}{2} region.  We show that five BIRS stars are possibly
pre-main sequence stars.  More infrared photometry, such as will be
available from 2MASS, is required to delineate the cluster properties.

\acknowledgments

We thank Henry Matthews for making the JCMT observations through the
CANSERV program.  C.R.K.  thanks the Ontario Graduate Scholarship
program for financial support during part of this study.  The research
of P.G.M. and D.J. is supported through grants from the Natural Sciences
and Engineering Research Council of Canada (NSERC).  D.R.B. acknowledges
financial support from the Commonwealth Scholarship and Fellowship Plan
and NSERC.



\clearpage

\figcaption[fig1.ps]{HIRES 60~$\mu$m image (linear
stretch from 0 -- 185~MJy/sr) of KR~140 with peaks corresponding to IRAS
point sources indicated by the white crosses. Contours show 1420~MHz
emission (contours at  5 -- 10~K at 1~K intervals). The IRAS point
source in the lower right corner of the image was not in the field of
view of our SCUBA observations.
 \label{fig:radioir}}

\figcaption[fig2.ps]{Jiggle maps of the NW part of the partial
dust shell surrounding KR~140 at 850~$\mu$m (top; linear stretch from
$-0.06$ to 0.1~Jy/beam) and 450~$\mu$m (bottom; linear stretch from
$-0.2$ to 0.05~Jy/beam; smoothed to 850~$\mu$m beam size).  Contours
of 12~$\mu$m emission (20 -- 32~MJy/sr at 2~MJy/sr intervals) show
that the 850~$\mu$m emission is coming from the same region as the
bright infrared emission.  The field of view is approximately $2'$ in each case
centered on $\alpha_{2000} = 2^{\rm h} 19^{\rm m} 44^{\rm s}$,
$\delta_{2000} = 61\degr 10\arcmin 45\arcsec$.   The hexagonal
shape is due to the arrangement of the bolometers in the detector
array. The location of these maps is indicated on the larger scan maps shown in
Figure~\ref{fig:scanmaps}.  
\label{fig:jigmap}}

\figcaption[fig3.ps]{Raw scan maps of KR~140 at 
850~$\mu$m (top; linear stretch from $-0.083$ to 0.151~Jy/beam)
and 450~$\mu$m (bottom; linear stretch from $-0.62$ to 0.86~Jy/beam).  
The locations of the jiggle maps shown in Figure~\ref{fig:jigmap} are 
indicated by the hexagons. 
\label{fig:scanmaps}}

\figcaption[fig4.ps]{Scan maps of KR~140, Weiner
filtered and with long wavelength modes suppressed. Top shows 850~$\mu$m
emission (linear stretch from $-0.045$ to 0.11~Jy/beam), and bottom shows
450~$\mu$m emission (linear stretch from $-0.30$ to
0.38~Jy/beam). Numbers and white crosses correspond to the source numbers
and positions listed in Table~\ref{tbl:sources}. 
\label{fig:scanmapsmooth}}

\figcaption[fig5.ps]{Scan map of KR~140 at 850~$\mu$m
(Figure~\ref{fig:scanmapsmooth}) compared with 1420 MHz emission (top
left, note central depression), HIRES 60~$\mu$m emission (top
right), and CO (J=1-0) emission (bottom). 
1420 MHz contours are from 5 -- 10~K at 1~K intervals and 60~$\mu$m
contours are from 100 -- 250~MJy/sr at 25~MJy/sr intervals.  
The bottom left figure shows integrated CO emission from $-45.5$
km~s$^{-1}$ to $-47.2$ km~s$^{-1}$ (V$_{\rm LSR}$); contours are from
2.5 -- 20~K at 2.5~K intervals. The bottom right figure shows
integrated CO emission from $-48.0$ km~s$^{-1}$ to $-53.7$ km~s$^{-1}$
(V$_{\rm LSR}$); note the doubling of scale with contours 5 -- 40~K at
5~K intervals. 
\label{fig:scancomp}}

\figcaption[fig6.ps]{Effect of changing temperature on the mass
derived using equation~\ref{eqn:scuba_mass}. Deviations in the derived
mass from starting temperatures of 10~K (solid line), 15~K (dashed
line), and 25~K (dash-dot line) are shown.  Horizontal solid lines
indicate changes by a factor of two.  At lower temperatures the derived
mass is much more sensitive to temperature.
\label{fig:terr}}

\figcaption[fig7.ps]{SED for the sources KMJB 15, 3, and 1, constructed
using cospatial HIRES and SCUBA data. The various symbols indicate measured
flux densities and the lines indicate the best-fitting bluebody curves
(fit to submillimeter far-infrared points only).  \label{fig:sourcesed}}

\figcaption[fig8.ps]{Closeup of the region surrounding IRAS
02168+6052. White crosses indicate KMJB 6 to 8 (top to bottom). The
star indicates the position of BIRS 128.
The greyscale image is the  850~$\mu$m scan map ($-0.04$
to 0.08 Jy/beam; linear stretch).  Solid contours are HIRES 60~$\mu$m
emission (130 -- 220 MJy/sr at 30 MJy/sr intervals).  Dashed contours
show integrated CO (J=1-0) emission (contours at 7, 9 and 11~K;
integrated from $-45.5$ km~s$^{-1}$ to $-47.2$ km~s$^{-1}$ (V$_{\rm
LSR}$)). \label{fig:eastshell}}

\figcaption[fig9.ps]{Closeup of the region
surrounding IRAS 02157+6053.  White crosses indicate KMJB 17 to 19
(l -- r).  The greyscale image is the 850~$\mu$m scan map linearly
stretched from $-0.028$ to 0.092 Jy/beam.  Contours of HIRES 60~$\mu$m
emission are overlayed (solid contours, 130--180 MJy/sr at 10 MJy/sr
intervals).  Integrated CO emission (V$_{\rm LSR} = -48.46$ to $-50.93$
km s$^{-1}$) is indicated by the dashed contours (6 -- 7.5~K at 0.5~K
intervals).
\label{fig:region5}}

\figcaption[fig10.ps]{Mass function for the submillimeter
sources in KR~140. The cumulative clump number is plotted against mass (jagged
solid line).  Overlaid are power-law distributions with different
indices, $0.49$ (best-fit; lower solid line), $1.0$ (dash-dot line), and
$1.5$ (steeper like typical stellar IMF; dash line).  
The $1.0$ and $1.5$ curves have been fixed to match the best-fit curve
at the third most massive object. 
\label{fig:clumps}}

\figcaption[fig11.ps]{Like Figure~\ref{fig:clumps} but for the large
cores in $\rho$~Oph \citep{mot98}.  Distributions with power-law indices 
$0.60$ (best-fit; solid lower line), $1.0$ (dash-dot) and $1.5$ (dash)
are shown. \label{fig:motte_clump}}

\figcaption[fig12.ps]{Mass determined from the submillimeter flux
density versus the Bonnor-Ebert mass ($M_{\rm BE}$) calculated for a sphere 
of the same radius and temperature.  The solid diagonal line (line of 
instability) represents the condition $M = M_{\rm BE}$.  Crosses indicate 
sources with $M>M_{\rm BE}$ and diamonds indicate those with $M<M_{\rm BE}$.
The asterisks represent the two sources known to contain embedded
stars.  Numbers correspond to the sources shown in
Figure~\ref{fig:scanmapsmooth}. \label{fig:massbe}}

\figcaption[fig13.ps]{Column density $M/D_{\rm eff}^{2}$
versus linear size.  The equivalent value of total visual extinction is
shown on the right y-axis. Note the bimodal distribution of sources.
 Symbols are the same as in Figure~\ref{fig:massbe}.  
\label{fig:massrad}}


\clearpage
\begin{deluxetable}{lccccccc}
\tablecaption{IRAS Point Sources Near KR~140 \label{tbl:iras}}
\tablewidth{0pt}
\tablehead{
                 & \colhead{$\alpha_{2000}$} & \colhead{$\delta_{2000}$} & \multicolumn{4}{c}{PSC Flux Density (Jy)} \\ \cline{4-7}
\colhead{IRAS} & \colhead{(h m s)}& \colhead{(\degr \phn \arcmin
\phn \arcsec)} & \colhead{12 $\mu$m} & \colhead{25 $\mu$m} &
\colhead{60 $\mu$m}  & \colhead{100 $\mu$m} &
\colhead{$\log\left(L_{\rm IR}/L_\odot \right)$\tablenotemark{a}}
}
\startdata
 02174+6052  & 02 21 08.5 & $+$61 06 00  & 0.88$\pm$0.05 & 2.46$\pm$0.1  & $\leq$32.0  & $\leq$127.9 & 2.8 \\
 02171+6058  & 02 20 51.4 & $+$61 12 01  & 0.36$\pm$0.05 & 1.8$\pm$0.1  & 11.6$\pm$1  & 63.5$\pm$11 & 2.5 \\
 02168+6052  & 02 20 33.0 & $+$61 05 56  & 2.2$\pm$0.5  & 2.2$\pm$0.5  & $\leq$32.0  & 127.9$\pm$18  & 2.9 \\
 02160+6057  & 02 19 47.3 & $+$61 11 24  & 2.4$\pm$0.4  & 3.0$\pm$0.6  & 47.4$\pm$9  & 215.1$\pm$34 & 3.0 \\
 02157+6053  & 02 19 24.6 & $+$61 07 15  & 0.82$\pm$0.13 & 1.31$\pm$0.17 & 21.8$\pm$3  & $\leq$215.1 & 2.9 \\
 02156+6045\tablenotemark{b}  & 02 19 21.2 & $+$60 59 45  &   $\leq$0.27  & 0.36$\pm$0.05 &  3.6$\pm$0.6 & $\leq$44.14 & 2.2\\
\enddata
\tablenotetext{a}{Calculated using \citet{eme88} technique and
assuming a distance of 2300 pc}
\tablenotetext{b}{Not in scanned region}
\end{deluxetable}

\clearpage
\begin{deluxetable}{lccccccccc}
\tablecaption{Submillimeter Sources\label{tbl:sources}}
\tablewidth{0pt}
\tablehead{ 
   \colhead{KMJB\tablenotemark{a}}  & \colhead{$\alpha_{2000}$} & \colhead{$\delta_{2000}$} & \colhead{$F_{450}$}  & \colhead{$F_{850}$} 
 & \colhead{$D_{\rm eff}$}    & \colhead{$T$}      & \colhead{$M$}            & \colhead{$\log(n)$}  & \colhead{$A_V$} \\
                     & \colhead{(h m s)} & \colhead{(\degr \phn \arcmin \phn \arcsec)} & \colhead{(Jy)} & \colhead{(Jy)} 
 & \colhead{(pc)} &  \colhead{(K)} & \colhead{($M_\odot$)}  & \colhead{cm$^{-3}$} & \colhead{}
}
\startdata
1   & 02 21 03.2 & $+$61 06 01    &  7.1$\pm$1     & 0.57$\pm$0.06 & 0.42    &  22\tablenotemark{b}  &  19  & 3.9    &  5.1    \\
2   & 02 20 59.2 & $+$61 07 12    &   \nodata      & 0.07$\pm$0.02 & 0.24    &  20\tablenotemark{c}  &   2  & 3.6   &  1.6    \\
3   & 02 20 51.2 & $+$61 11 50    &   \nodata      & 0.28$\pm$0.01 & 0.23    &  27\tablenotemark{b}  &   5  & 4.1   &  4.5    \\
4   & 02 20 52.0 & $+$61 08 33    &  0.5$\pm$0.2   & 0.20$\pm$0.05 & 0.35    &   7\tablenotemark{d}  &  40  & 4.4   &  15.4   \\
5   & 02 20 41.4 & $+$61 09 54    &  0.82$\pm$0.2  & 0.15$\pm$0.02 & 0.55    &  12\tablenotemark{d}  &   9  & 3.2   &  1.4    \\
6   & 02 20 35.3 & $+$61 07 40    &   \nodata      & 0.28$\pm$0.03 & 0.45    &   7\tablenotemark{d}  &  56  & 4.3   &  13.1   \\  
6a  & 02 20 35.0 & $+$61 08 35    &  0.43$\pm$0.1  &   \nodata     & 0.37    &   7\tablenotemark{d}  &  33  & 4.3   &  10.8   \\
6b  & 02 20 33.2 & $+$61 08 01    &  0.02$\pm$0.01 &   \nodata     & $<0.10$ &   7\tablenotemark{d}  &   2  & $>4.6$ &  $>7.3$ \\
6c  & 02 20 36.3 & $+$61 07 26    &  0.24$\pm$0.1  &   \nodata     & 0.48    &   7\tablenotemark{d}  &  19  & 3.7    &  3.5    \\
7   & 02 20 35.2 & $+$61 06 13    &   \nodata      & 0.07$\pm$0.02 & $<0.18$ &  30\tablenotemark{c}  &   1  & $>3.7$ &  $>1.5$ \\
8   & 02 20 30.4 & $+$61 05 05    &  2.2$\pm$0.1   & 0.65$\pm$0.06 & 0.56    &   8\tablenotemark{d}  &  80  & 4.1    &  12.0   \\ 
9   & 02 20 06.3 & $+$61 01 21    &  0.40$\pm$0.05 & 0.28$\pm$0.03 & 0.48    &   7\tablenotemark{d}  &  77  & 4.3    &  15.8   \\
10  & 02 20 06.5 & $+$61 12 10    &   \nodata      & 0.06$\pm$0.01 & 0.36    &  20\tablenotemark{c}  &   2  & 3.1    &  0.7    \\
11  & 02 20 04.0 & $+$61 04 30    &  1.0$\pm$0.5   & 0.17$\pm$0.02 & 0.26    &  13\tablenotemark{d}  &   9  & 4.2    &  6.3    \\
12  & 02 19 55.0 & $+$61 05 13    &  1.0$\pm$0.3   & 0.50$\pm$0.05 & 0.30    &   7\tablenotemark{d}  & 130  & 5.1    &  68.2   \\
13  & 02 19 54.0 & $+$61 02 05    &   \nodata      & 0.42$\pm$0.02 & 0.61    &  20\tablenotemark{c}  &  11  & 3.1    &  1.4    \\
14  & 02 19 45.5 & $+$61 05 45    &  0.68$\pm$0.2  & 0.17$\pm$0.04 & 0.23    &   9\tablenotemark{d}  &  17  & 4.6    &  15.2   \\
15  & 02 19 43.3 & $+$61 11 27    &  4.1$\pm$0.9   & 0.8$\pm$0.2   & 0.69    &  28\tablenotemark{b}  &   9  & 2.9    &  0.9    \\
16  & 02 19 46.7 & $+$61 01 33    &   \nodata      & 0.11$\pm$0.02 & 0.36    &  20\tablenotemark{c}  &   3  & 3.3    &  1.1    \\
17  & 02 19 33.3 & $+$61 07 44    &   \nodata      & 0.22$\pm$0.07 & $<0.18$ &  20\tablenotemark{c}  &   6  & $>4.5$ &  8.7    \\
18  & 02 19 29.9 & $+$61 07 26    &   \nodata      & 0.16$\pm$0.05 & $<0.18$ &  20\tablenotemark{c}  &   4  & $>4.3$ &  5.8    \\
19  & 02 19 22.7 & $+$61 07 02    &  1.27$\pm$0.3  & 0.39$\pm$0.04 & 0.33    &   8\tablenotemark{d}  &  50  & 4.6    &  21.7   \\
20  & 02 19 11.7 & $+$61 06 06    &  \nodata       & 0.09$\pm$0.01 & 0.25    &  20\tablenotemark{c}  &   2  & 3.3    &  1.5    \\
21  & 02 19 54.4 & $+$61 01 56    & 0.26$\pm$0.05  & \nodata       & 0.26    &  20\tablenotemark{c}  & 0.8  & 3.1    &  0.6    \\
22  & 02 19 49.1 & $+$61 01 43    & 0.15$\pm$0.06  & \nodata       & 0.23    &  20\tablenotemark{c}  & 0.7  & 3.2    &  0.6    \\
\enddata
\tablenotetext{a}{Acronym for the new submillimeter sources; from Kerton,
Martin, Johnstone, \& Ballantyne}
\tablenotetext{b}{$T$ from submm-IR SED fit}
\tablenotetext{c}{$T$ assumed}
\tablenotetext{d}{$T$ from 450/850~$\mu$m color}
\end{deluxetable}


\clearpage
\begin{deluxetable}{lll}
\tablecaption{Dust Opacity at 850~$\mu$m  \label{tbl:cvals}}
\tablewidth{0pt}
\tablehead{
\colhead{$C$ (g cm$^{-2}$)\tablenotemark{a}}  & \colhead{Source}  & \colhead{Recommended Use} 
}
\startdata
286  &  \citet{dra90} \& \citet{dra84}  & diffuse ISM \\
116  &  \citet{hil83}                   & diffuse ISM, canonical value \\
85.5 &  \citet{pre93}			& dense clumps and cores \\
53.4 &  \citet{oss94} 			& regions with $n \approx 10^5$ cm$^{3}$ \\
38.9 &  \citet{oss94}                   & regions with $n \geq 10^7$ cm$^{3}$ \\
21.4 &  \citet{kru94}			& dense, cold regions \\
\enddata
\tablenotetext{a}{$C$ combines the dust opacity ($\kappa_\nu$) and
the dust to gas mass ratio ($C = M_g/(M_d \kappa_\nu)$)}
\end{deluxetable}

\end{document}